\documentclass[manuscript,screen]{acmart}
\AtBeginDocument{%
  }

\setcopyright{acmlicensed}
\copyrightyear{2026}
\acmYear{2026}
\acmDOI{XXXXXX}
\acmConference{CHI 2026 Workshop Herding CATs: Making Sense of Creative Activity Traces}{April 15, 2026}{Barcelona, Spain}

\begin{document}

\title{Chasing RATs: Tracing Reading for and as Creative Activity}

\author{Sophia Liu}
\email{sophiawliu@berkeley.edu}
\orcid{0009-0008-7746-0749}
\affiliation{%
  \institution{University of California, Berkeley}
  \city{Berkeley}
  \state{California}
  \country{USA}
}

\author{Shm Garanganao Almeda}
\email{shm.almeda@berkeley.edu}
\orcid{0000-0001-7660-313X}
\affiliation{%
  \institution{University of California, Berkeley}
  \city{Berkeley}
  \state{California}
  \country{USA}
}

\renewcommand{\shortauthors}{Liu \& Almeda}

\begin{abstract}
Creativity research has privileged making over the interpretive labor that precedes and shapes it. We introduce Reading Activity Traces (RATs), a proposal that treats reading---broadly defined to include navigating, interpreting, and curating media across interconnected sources---as creative activity both \textit{for} future artifacts and \textit{as} a form of creation in its own right. By tracing trajectories of traversal, association, and reflection as inspectable artifacts, RATs render visible the creative work that algorithmic feeds and AI summarization increasingly compress and automate away. We illustrate this through WikiRAT, a speculative instantiation on Wikipedia, and open new ground for reflective practice, reader modeling, collective sensemaking, and understanding what is lost when human interpretation is automated---towards designing intelligent tools that preserve it.
\end{abstract}



\received{20 February 2026}


\maketitle

\section{Introduction}
\begin{quote}
    \textit{Our literature is characterized by the pitiless divorce which the literary institution maintains between the producer of the text and its user, between its owner and its customer, between its author and its reader. This reader is thereby plunged into a kind of idleness...} \begin{flushright}--- Roland Barthes\end{flushright}
\end{quote}
Creative work rarely begins with artifact production; it begins with the labor of finding, gathering, and interpreting what already exists. Writers, designers, and researchers all situate themselves within prior work to form the conceptual models that precede drafting or sketching. These activities shape internal models even when no visible artifact is produced. To fully support the creative process, we must account for these earlier phases of discovery and sensemaking.

Shneiderman's account of Creativity Support Tools (CSTs) explicitly includes discovery and exploratory search as creative activities~\cite{Shneiderman2007}. Yet CSTs tend to center \textit{composition}, neglecting the position of artifact production within wider creative ecosystems that depend upon media distribution and reception activities~\cite{Almeda2025, Chung2022}. If we aim to trace creative activity, we should consider how reading shapes what creators produce, and how interpretation itself can be a meaningful and generative creative act.

We extend Creative Activity Traces (CATs) into \textbf{Reading Activity Traces (RATs)}---using \textit{reading} to broadly describe navigating, interpreting, and curating media---especially across interconnected sources, as afforded by the web's hypermedia structures. Reading operates \textit{\textbf{for}} creative activity when it inspires future artifacts, and \textit{\textbf{as}} creative activity when navigation itself constitutes a reader-led trajectory of sensemaking. Hypertextual traversal, such as ``Wikipedia rabbit-holing,'' exemplifies reading as a creative act: users author knowledge paths by following links, even in the absence of visible artifact production~\cite{Amisola2024}.

We propose RATs as a way of characterizing reading as creative activity, and illustrate the concept in WikiRAT, a speculative instantiation on Wikipedia. We situate this proposal in creativity and sensemaking research, framing reading traces as a design material that can broaden how we understand and support creative processes~\cite{Hammad2025, Li2023}.

\section{Related Work}
Many media and cultural domains recognize reading as an active process. Aarseth’s \textit{ergodic literature} frames traversal as non-trivial work where meaning emerges through a reader’s effortful movement across a textual system \cite{Aarseth1997}. The work of internet artist Chia Amisola recognizes web browsing and hypertextual navigation as a performance of authorship and creativity~\cite{Amisola2024}.
In HCI, Bardzell's interaction criticism centers aesthetic interpretation in the evaluation of systems as cultural artifacts, regarding the user as an interpretative Reader/Viewer~\cite{Bardzell2011}. Critical Creativity Support Tool (CST) research has noted a reductive overemphasis on designing and evaluating CSTs for productivity and artifact-production~\cite{Li2023, RhysCox2025}, and have called for methodological shifts towards understanding creativity support, especially in the context of \textit{artist} support, as socially and culturally entangled~\cite{Chung2022, Li2023, Almeda2025, Kato2025}. This ``artifact bias'' in CST research has left the sensemaking layer of creative research neglected, as what some have called ``creative dark matter''---meaningful creative and intellectual work that remains invisible \cite{Sholette2004, Tamari2024, Chan2021}.

Valuing the process of reading shifts focus from creative products toward creative experiences, echoing prior work recognizing dialectical~\cite{Zhang2024} and autotelic~\cite{Compton2019} creative activities---where value is intrinsic to the activity itself, rather than its outcomes. Kreminski \& Mateas' ``reflective creators'' extend autotelic creativity support, offering \textit{process aesthetics} as a lens for analytically interpreting the experiential qualities of a CST~\cite{Kreminski2021}. CATs represent a shift toward valuing and understanding process, with techniques like sensemaking curves \cite{Davis2017} and fuzzy linkography \cite{Smith2025}. RATs extend this turn towards creative interpretation and reception---as one example, Artographer used traces of user activity to characterize how users engaged with artwork within a spatial art presentation interface~\cite{AlmedaArtographer2025}.

While digital tools for supporting artifact creation (e.g., for \textit{making} images, video, music) abound, the space of systems for \textit{reading} media is increasingly dominated by opaque, monolithic, algorithmically-mediated social media platforms that embed the values of the corporations governing them \cite{feng_mapping_2024, von_davier_looking_2025}. AI-generated summarization and ``infinite-scroll'' content feeds restructure the experience of search and browsing, privileging passive consumption over user-directed exploration \cite{baughan_i_2022}. This compression stands in tension with processes where friction, slowness, and seamfulness are recognized as positive and desirable design values \cite{Kreminski2021, Ehsan2024, Liu2025, AlmedaArtographer2025}.

Emerging content discovery systems challenge this paradigm by treating reading as a generative act grounded in association---the intentional following of links and the contextual understanding of one idea in relation to another. This associative sensemaking is increasingly under pressure as LLM-mediated dialogue and algorithmic recommendation displace the reader's own acts of finding and selecting \cite{Rahdari2025}. Are.na and Sublime offer graph-like environments where users find inspiration outside algorithmic feeds, foregrounding associative links as creative output and making those links shareable and buildable-upon by others \cite{AreNa, Sublime}. In scientific communication, Discourse Graphs and Semble externalize the relational labor of reading, formalizing the associative layer as part of the intellectual record and empowering collective sensemaking~\cite{Chan2021, Tamari2024}.

Through RATs, we might seek to better understand and support reading---navigation, interpretation, and curation---as a creative epistemic process~\cite{Tricaud2024}, and to empower designers to move against hegemonic narratives that privilege outputs at the expense of process. We see RATs as a way to make readers' associative paths legible and shareable, and to open new questions for CST researchers and designers about what it means to support creativity beyond production.

\section{WikiRAT}

To illustrate and explore the concept of RATs, we propose WikiRAT, a speculative browser extension for tracing reading activity on Wikipedia. WikiRAT reifies reading, \textit{for} and \textit{as} creative activity, as a structured trace.

A reader on Wikipedia begins on the article for ``Intentional community,'' following hyperlinks from the ``Whole Earth Catalog'' to ``Stewart Brand,'' then to ``Doug Engelbart,'' ``Hypertext,'' and finally the ``World Wide Web.'' Along the way, article titles reappear as hyperlinks embedded into text on new pages, prompting recognition as implicit associations form across contexts. The reader may revisit earlier pages as concepts resurface in new configurations. This non-linear movement reflects a core creative affordance of hypertext. 
The reader may later use the associative knowledge they've gained \textit{for} making, e.g., to produce a research paper or write a blog post. They may also simply close the tab. In either case, creative work has already unfolded. The sequence of interpretation, association, and navigational choices constitutes reading \textit{as} creative activity. WikiRAT traces this reader-driven path, making implicit connections explicit, and embedding the creative and associative knowledge produced in this process as a design material.

Wikipedia is a well-suited site for this exploration, as its hyperlink topology is explicit and machine-readable. Prior work has shown that Wikipedia rabbit hole sessions exhibit traceable and measurable structure \cite{Piccardi2022}. Unlike algorithmically-curated feeds and AI summaries that automate and compress the act of reading, Wikipedia readily facilitates user-driven knowledge and media traversal.

We are interested in designing WikiRAT to trace reading at three levels, and producing a representation for each:

\begin{enumerate}
    \item A \textbf{hyperlinkograph}\footnote{We borrow the term \textit{linkograph} from design process research, where it denotes a graph of links between creative moves \cite{Goldschmidt2014}. We repurpose it here for reading moves. The \textit{hyperlinkograph} takes inspiration from prior work reconstructing hypertextual navigation on Wikipedia as trees~\cite{Piccardi2022}.} records the directed path of hyperlinks taken by the reader, annotated with engagement signals such as dwell time.
    
    \item A computational \textbf{fuzzy linkograph}~\cite{Smith2025} adds edges between semantically similar articles, surfacing conceptual relationships the reader traversed implicitly, beyond explicitly clicked links. 
    
    \item A \textbf{reflective linkograph} invites readers to manually draw connections between concepts that stuck with them from their traversal, towards understanding reflection, retrospection, and reinterpretation as creative acts \cite{Kreminski2021}.
\end{enumerate}

By tracing navigation, association, and retrospection, WikiRAT seeks to help us understand reading and interpretation as a layered, sociotechnically entangled activity. The next section considers the implications of tracing reading in this way and explores what WikiRAT makes possible.

\section{Implications \& Future Work}

We are interested in how WikiRAT can generate and present reading traces for a variety of audiences and uses; here we outline three: individual readers, interpretive communities, and researchers, especially in the context of Human-AI creativity research.

For individual readers, the trace becomes a reflective epistemic artifact~\cite{Tricaud2024}, a map of their own sensemaking that can be revisited and reused. For creativity researchers, WikiRAT opens empirical ground for comparison across readers, comparison across the three linkograph types, and reader modeling that extends aggregate analysis \cite{Piccardi2022} to individual, inspectable trajectories \cite{Alvarez2022}. For system designers, WikiRAT data could inform adaptive systems where recommendations arise from a reader's own traversal history rather than global popularity metrics \cite{Gonzalez2019}, and because the model is derived from a legible trace, it can remain inspectable and adjustable by the reader, allowing them to see and reshape the signals that inform adaptation rather than having their exploratory labor optimized away \cite{Chapekis2025}.

Aggregated RATs are a promising material for designing collective and social experiences. Collective traces externalize the interpretive labor of many readers as an inspectable, navigable artifact~\cite{Tamari2024, Chan2021}. Individual traces, aggregated across many readers and sessions, can become shared infrastructure, surfacing desire paths that are well-traveled, versus areas that remain uncharted by the community. This positions RATs as a human-community-driven alternative to algorithmic-curation in the design of social media platforms. Platforms like Are.na and Discourse Graphs already gesture toward this model for curation and scholarly sensemaking \cite{AreNa, DiscourseGraphs}; as a framework for CST research and development, we propose RATs as a move towards more deeply understanding, and designing for, the ways communities learn and make meaning together.  

As AI browsing agents and summarization tools increasingly perform research on behalf of users \cite{Yang2025, Chapekis2025, White2024}, the dialectical, temporally unfolding labor of reading-it-yourself becomes harder to justify and easier to outsource. We propose RATs as a necessary tool for addressing an urgent and timely space of questions. How does a human-driven rabbit hole differ from that of an AI agent? What do we stand to lose in the automation of interpretive processes? Can an AI agent be curious and creative? Emerging evidence suggests that AI summaries reduce click-through, shorten browsing sessions, and redirect attention away from cited sources, compressing what might have unfolded as dynamic traversal into a single static output~\cite{Chapekis2025}. Comparing human and AI-agent reading traces can help make legible what is retained, and what is lost, in automation---in navigational structure, associative logic, dwell behavior, and in process aesthetics, e.g., the moments of surprise that drive a human reader to \textit{draw a link} between concepts. RATs can provide the instrumentation to surface and study these differences at depth.

This paper is a position and proposal. Our immediate future work involves implementing WikiRAT (which we imagine as a browser extension or experimental substrate) and instrumenting it in a user study to trace and study reading activities. This involves considering which reading activity signals are tractable to log and meaningful to analyze, and how to best organize and visualize this information, starting with the representations and use cases proposed above. Beyond Wikipedia, the RAT framework extends naturally to browsing more broadly (Browsing Activity Traces, or BATs), informational retrieval, and any information system where \textit{humans} are conducting interpretive labor.

\section{Conclusion}
RATs extend CATs into interpretation \textit{for} and \textit{as} creative activity. By making knowledge and media traversal legible, we might embed---and preserve---the work of sensemaking that contemporary information system design paradigms are increasingly compressing, automating, and obscuring. As AI-mediation reshapes the ways we find and interpret knowledge, we hope that RATs can empower readers as creative actors, and provide design material for counter-hegemony.

\bibliographystyle{ACM-Reference-Format}
\bibliography{reference}

@inproceedings{Almeda2025,
author = {Almeda, Shm Garanganao and Kim, Joy O and Hartmann, Bjoern},
title = {Creativity Supportive Ecosystems: A Framework for Understanding Function and Disruption in Online Art Worlds},
year = {2025},
isbn = {9798400713941},
publisher = {Association for Computing Machinery},
address = {New York, NY, USA},
url = {https://doi.org/10.1145/3706598.3713734},
doi = {10.1145/3706598.3713734},
booktitle = {Proceedings of the 2025 CHI Conference on Human Factors in Computing Systems},
articleno = {267},
numpages = {17},
location = {
},
series = {CHI '25}
}

@ARTICLE{Alvarez2022,
  author={Alvarez, Alberto and Font, Jose and Togelius, Julian},
  journal={IEEE Transactions on Games}, 
  title={Toward Designer Modeling Through Design Style Clustering}, 
  year={2022},
  volume={14},
  number={4},
  pages={676-686},
  doi={10.1109/TG.2022.3143800}}

@inproceedings{Smith2025,
author = {Smith, Amy and Anderson, Barrett R and Otto, Jasmine Tan and Karth, Isaac and Sun, Yuqian and Joon Young Chung, John and Roemmele, Melissa and Kreminski, Max},
title = {Fuzzy Linkography: Automatic Graphical Summarization of Creative Activity Traces},
year = {2025},
isbn = {9798400712890},
publisher = {Association for Computing Machinery},
address = {New York, NY, USA},
url = {https://doi.org/10.1145/3698061.3726915},
doi = {10.1145/3698061.3726915},
booktitle = {Proceedings of the 2025 Conference on Creativity and Cognition},
pages = {637–650},
numpages = {14},
location = {
},
series = {C\&C '25}
}

@inproceedings{Davis2017,
author = {Davis, Nicholas and Hsiao, Chih-Pin and Singh, Kunwar Yashraj and Lin, Brenda and Magerko, Brian},
title = {Creative Sense-Making: Quantifying Interaction Dynamics in Co-Creation},
year = {2017},
isbn = {9781450344036},
publisher = {Association for Computing Machinery},
address = {New York, NY, USA},
url = {https://doi.org/10.1145/3059454.3059478},
doi = {10.1145/3059454.3059478},
booktitle = {Proceedings of the 2017 ACM SIGCHI Conference on Creativity and Cognition},
pages = {356–366},
numpages = {11},
location = {Singapore, Singapore},
series = {C\&C '17}
}

@article{Shneiderman2007,
author = {Shneiderman, Ben},
title = {Creativity support tools: accelerating discovery and innovation},
year = {2007},
issue_date = {December 2007},
publisher = {Association for Computing Machinery},
address = {New York, NY, USA},
volume = {50},
number = {12},
issn = {0001-0782},
url = {https://doi.org/10.1145/1323688.1323689},
doi = {10.1145/1323688.1323689},
journal = {Commun. ACM},
month = dec,
pages = {20–32},
numpages = {13}
}

@misc{Amisola2024,
	title = {Becoming hypertext},
	url = {https://everythingi.love},
	language = {en},
	urldate = {2025-09-10},
	journal = {Everything I Love {\textbar} Chia Amisola},
	author = {Amisola, Chia},
	month = dec,
	year = {2024},
}

@article{Bardzell2011,
    author = {Bardzell, Jeffrey},
    title = {Interaction criticism: An introduction to the practice},
    journal = {Interacting with Computers},
    volume = {23},
    number = {6},
    pages = {604-621},
    year = {2011},
    month = {07},
    issn = {0953-5438},
    doi = {10.1016/j.intcom.2011.07.001},
    url = {https://doi.org/10.1016/j.intcom.2011.07.001},
    eprint = {https://academic.oup.com/iwc/article-pdf/23/6/604/2284024/iwc23-0604.pdf},
}

@phdthesis{Compton2019,
author={Compton,Katherine},
year={2019},
title={Casual Creators: Defining a Genre of Autotelic Creativity Support Systems},
journal={ProQuest Dissertations and Theses},
pages={685},
note={Copyright - Database copyright ProQuest LLC; ProQuest does not claim copyright in the individual underlying works; Last updated - 2023-06-21},
isbn={9781085668880},
language={English},
url={https://www.proquest.com/dissertations-theses/casual-creators-defining-genre-autotelic/docview/2300563742/se-2},
}

@inproceedings{Kreminski2021,
  title={Reflective Creators},
  author={Max Kreminski and Michael Mateas},
  booktitle={ICCC},
  year={2021},
  url={https://api.semanticscholar.org/CorpusID:237181869}
}

@inproceedings{Zhang2024,
author = {Zhang, Haoqi},
title = {Searching for the Non-Consequential: Dialectical Activities in HCI and the Limits of Computers},
year = {2024},
isbn = {9798400703300},
publisher = {Association for Computing Machinery},
address = {New York, NY, USA},
url = {https://doi.org/10.1145/3613904.3641945},
doi = {10.1145/3613904.3641945},
booktitle = {Proceedings of the 2024 CHI Conference on Human Factors in Computing Systems},
articleno = {297},
numpages = {13},
location = {Honolulu, HI, USA},
series = {CHI '24}
}

@book{Aarseth1997,
  title={Cybertext: Perspectives on Ergodic Literature},
  author={Aarseth, Espen J.},
  year={1997},
  publisher={Johns Hopkins University Press},
  address={Baltimore, Md.},
  isbn={978-0801855795}
}

@article{Chan2021,
	author = {Chan, Joel},
	journal = {Commonplace},
	number = {1},
	year = {2021},
	month = {aug 23},
	note = {https://commonplace.knowledgefutures.org/pub/m76tk163},
	publisher = {},
	title = {Sustainable {Authorship} {Models} for a {Discourse}-{Based} {Scholarly} {Communication} {Infrastructure}},
	volume = {1},
}

@online{AreNa,
  author    = {{Are.na}},
  title     = {Are.na},
  year      = {n.d.},
  url       = {https://www.are.na},
  urldate   = {2026-02-20}
}

@online{Sublime,
  author    = {{Sublime}},
  title     = {Sublime},
  year      = {n.d.},
  url       = {https://sublime.app},
  urldate   = {2026-02-20}
}

@online{DiscourseGraphs,
  author    = {{Discourse Graphs}},
  title     = {Discourse Graphs},
  year      = {n.d.},
  url       = {https://discoursegraphs.com},
  urldate   = {2026-02-20}
}

@article{Tamari2024,
  title={Sensemaking Networks: Transforming Social Media into a Sensemaking Layer for Science},
    year      = {2024},
  author={Tamari, Ronen and Ospina, Pepo and Oriel, Shahar and Finck, Wesley},
}

@inproceedings{Liu2025,
author = {Liu, Sophia and Almeda, Shm Garanganao},
title = {Agency Among Agents: Designing with Hypertextual Friction in the Algorithmic Web},
year = {2025},
isbn = {9798400715334},
publisher = {Association for Computing Machinery},
address = {New York, NY, USA},
url = {https://doi.org/10.1145/3720533.3750065},
doi = {10.1145/3720533.3750065},
booktitle = {Adjunct Proceedings of the 36th ACM Conference on Hypertext and Social Media},
pages = {30–34},
numpages = {5},
location = {
},
series = {HT Adjunct '25}
}

@article{Sholette2004,
  title={Dark Matter: Activist Art and the Counter-Public Sphere},
  author={Sholette, Gregory},
  journal={Journal of Aesthetics and Protest},
  volume={3},
  year={2004},
  url={https://www.darkmatterarchives.net/wp-content/uploads/2012/01/05_darkmatter.pdf}
}

@misc{AlmedaArtographer2025,
      title={Artographer: a Curatorial Interface for Art Space Exploration}, 
      author={Shm Garanganao Almeda and John Joon Young Chung and Bjoern Hartmann and Sophia Liu and Brett Halperin and Yuwen Lu and Max Kreminski},
      year={2025},
      eprint={2512.02288},
      archivePrefix={arXiv},
      primaryClass={cs.HC},
      url={https://arxiv.org/abs/2512.02288}, 
}

@article{von_davier_looking_2025,
	title = {Looking for {Art} in a {Sea} of {Content}: {A} {Human}-{Centered} {Approach} to {Supporting} {Creativity} on {Social} {Media}},
	volume = {9},
	shorttitle = {Looking for {Art} in a {Sea} of {Content}},
	url = {https://dl.acm.org/doi/10.1145/3711025},
	doi = {10.1145/3711025},
	number = {2},
	urldate = {2026-01-30},
	journal = {Proc. ACM Hum.-Comput. Interact.},
	author = {von Davier, Thomas Serban and Noh, Hayoun and Van Kleek, Max and Shadbolt, Nigel},
	month = may,
	year = {2025},
	pages = {CSCW127:1--CSCW127:25},
}

@inproceedings{baughan_i_2022,
	address = {New York, NY, USA},
	series = {{CHI} '22},
	title = {“{I} {Don}’t {Even} {Remember} {What} {I} {Read}”: {How} {Design} {Influences} {Dissociation} on {Social} {Media}},
	isbn = {978-1-4503-9157-3},
	shorttitle = {“{I} {Don}’t {Even} {Remember} {What} {I} {Read}”},
	url = {https://dl.acm.org/doi/10.1145/3491102.3501899},
	doi = {10.1145/3491102.3501899},
	urldate = {2026-02-05},
	booktitle = {Proceedings of the 2022 {CHI} {Conference} on {Human} {Factors} in {Computing} {Systems}},
	publisher = {Association for Computing Machinery},
	author = {Baughan, Amanda and Zhang, Mingrui Ray and Rao, Raveena and Lukoff, Kai and Schaadhardt, Anastasia and Butler, Lisa D. and Hiniker, Alexis},
	month = apr,
	year = {2022},
	pages = {1--13},
}

@inproceedings{feng_mapping_2024,
	address = {New York, NY, USA},
	series = {{CHI} '24},
	title = {Mapping the {Design} {Space} of {Teachable} {Social} {Media} {Feed} {Experiences}},
	isbn = {979-8-4007-0330-0},
	url = {https://dl.acm.org/doi/10.1145/3613904.3642120},
	doi = {10.1145/3613904.3642120},
	urldate = {2026-02-05},
	booktitle = {Proceedings of the 2024 {CHI} {Conference} on {Human} {Factors} in {Computing} {Systems}},
	publisher = {Association for Computing Machinery},
	author = {Feng, K. J. Kevin and Koo, Xander and Tan, Lawrence and Bruckman, Amy and McDonald, David W. and Zhang, Amy X.},
	month = may,
	year = {2024},
	pages = {1--20},
}

@inproceedings{Hammad2025,
author = {Hammad, Noor},
title = {Towards Trail-Aware Digital Content Creation Tools},
year = {2025},
isbn = {9798400720239},
publisher = {Association for Computing Machinery},
address = {New York, NY, USA},
url = {https://doi.org/10.1145/3744736.3749328},
doi = {10.1145/3744736.3749328},
booktitle = {Companion Proceedings of the Annual Symposium on Computer-Human Interaction in Play},
pages = {273–275},
numpages = {3},
location = {
},
series = {CHI PLAY Companion '25}
}

@inproceedings{Piccardi2022,
author = {Piccardi, Tiziano and Gerlach, Martin and West, Robert},
title = {Going Down the Rabbit Hole: Characterizing the Long Tail of Wikipedia Reading Sessions},
year = {2022},
isbn = {9781450391306},
publisher = {Association for Computing Machinery},
address = {New York, NY, USA},
url = {https://doi.org/10.1145/3487553.3524930},
doi = {10.1145/3487553.3524930},
booktitle = {Companion Proceedings of the Web Conference 2022},
pages = {1324–1330},
numpages = {7},
location = {Virtual Event, Lyon, France},
series = {WWW '22}
}

@book{Goldschmidt2014,
  author    = {Goldschmidt, Gabriela},
  title     = {Linkography: Unfolding the Design Process},
  publisher = {The MIT Press},
  year      = {2014},
  doi       = {10.7551/mitpress/9455.001.0001},
  isbn      = {9780262322157},
  address   = {Cambridge, MA}
}

@misc{Yang2025,
      title={Agentic Web: Weaving the Next Web with AI Agents}, 
      author={Yingxuan Yang and Mulei Ma and Yuxuan Huang and Huacan Chai and Chenyu Gong and Haoran Geng and Yuanjian Zhou and Ying Wen and Meng Fang and Muhao Chen and Shangding Gu and Ming Jin and Costas Spanos and Yang Yang and Pieter Abbeel and Dawn Song and Weinan Zhang and Jun Wang},
      year={2025},
      eprint={2507.21206},
      archivePrefix={arXiv},
      primaryClass={cs.AI},
      url={https://arxiv.org/abs/2507.21206}, 
}

@misc{Chapekis2025,
  author       = {Chapekis, Athena and Lieb, Anna},
  title        = {Google users are less likely to click on links when an {AI} summary appears in the results},
  howpublished = {Pew Research Center},
  year         = {2025},
  month        = {July},
  url          = {https://www.pewresearch.org/short-reads/2025/07/22/google-users-are-less-likely-to-click-on-links-when-an-ai-summary-appears-in-the-results/},
  note         = {Accessed: 2026-02-21}
}

@article{White2024,
author = {White, Ryen W.},
title = {Advancing the Search Frontier with AI Agents},
year = {2024},
issue_date = {September 2024},
publisher = {Association for Computing Machinery},
address = {New York, NY, USA},
volume = {67},
number = {9},
issn = {0001-0782},
url = {https://doi.org/10.1145/3655615},
doi = {10.1145/3655615},
journal = {Commun. ACM},
month = aug,
pages = {54–65},
numpages = {12}
}

@inproceedings{Li2023,
author = {Li, Jingyi and Rawn, Eric and Ritchie, Jacob and Tran O'Leary, Jasper and Follmer, Sean},
title = {Beyond the Artifact: Power as a Lens for Creativity Support Tools},
year = {2023},
isbn = {9798400701320},
publisher = {Association for Computing Machinery},
address = {New York, NY, USA},
url = {https://doi.org/10.1145/3586183.3606831},
doi = {10.1145/3586183.3606831},
booktitle = {Proceedings of the 36th Annual ACM Symposium on User Interface Software and Technology},
articleno = {47},
numpages = {15},
location = {San Francisco, CA, USA},
series = {UIST '23}
}

@INPROCEEDINGS{Gonzalez2019,
  author={González, Daniel and Tansini, Libertad},
  booktitle={2019 XLV Latin American Computing Conference (CLEI)}, 
  title={Modelling Traceability in Recommender Systems}, 
  year={2019},
  volume={},
  number={},
  pages={1-6},
  doi={10.1109/CLEI47609.2019.235091}}

@inproceedings{RhysCox2025,
author = {Rhys Cox, Samuel and B\o{}jer Djern\ae{}s, Helena and van Berkel, Niels},
title = {Beyond Productivity: Rethinking the Impact of Creativity Support Tools},
year = {2025},
isbn = {9798400712890},
publisher = {Association for Computing Machinery},
address = {New York, NY, USA},
url = {https://doi.org/10.1145/3698061.3726924},
doi = {10.1145/3698061.3726924},
booktitle = {Proceedings of the 2025 Conference on Creativity and Cognition},
pages = {735–749},
numpages = {15},
location = {
},
series = {C\&C '25}
}

@inproceedings{Kato2025,
  author    = {Jun Kato and Hiromu Yakura},
  title     = {Power, Culture, and Sustainability in Creativity Support Tools: A Post-growth Perspective},
  booktitle = {Hybrid Workshop: Advancing Post-growth HCI at CHI '25},
  year      = {2025},
  month     = {April},
  address   = {Yokohama, Japan and online},
  pages     = {2},
  note      = {April 26, 2025},
  institution = {AIST, Japan and Université Paris-Saclay, France and Max-Planck Institute for Human Development, Germany}
}

@inproceedings{Chung2022,
author = {Chung, John Joon Young and He, Shiqing and Adar, Eytan},
title = {Artist Support Networks: Implications for Future Creativity Support Tools},
year = {2022},
isbn = {9781450393584},
publisher = {Association for Computing Machinery},
address = {New York, NY, USA},
url = {https://doi.org/10.1145/3532106.3533505},
doi = {10.1145/3532106.3533505},
booktitle = {Proceedings of the 2022 ACM Designing Interactive Systems Conference},
pages = {232–246},
numpages = {15},
location = {Virtual Event, Australia},
series = {DIS '22}
}

@article{Ehsan2024,
author = {Ehsan, Upol and Liao, Q. Vera and Passi, Samir and Riedl, Mark O. and Daum\'{e}, Hal, III},
title = {Seamful XAI: Operationalizing Seamful Design in Explainable AI},
year = {2024},
issue_date = {April 2024},
publisher = {Association for Computing Machinery},
address = {New York, NY, USA},
volume = {8},
number = {CSCW1},
url = {https://doi.org/10.1145/3637396},
doi = {10.1145/3637396},
journal = {Proc. ACM Hum.-Comput. Interact.},
month = apr,
articleno = {119},
numpages = {29}
}

@inproceedings{Rahdari2025,
author = {Rahdari, Behnam and Brusilovsky, Peter},
title = {From Links to Dialogue; Hypertext Challenges and Opportunities in Conversational Navigation},
year = {2025},
isbn = {9798400715334},
publisher = {Association for Computing Machinery},
address = {New York, NY, USA},
url = {https://doi.org/10.1145/3720533.3750064},
doi = {10.1145/3720533.3750064},
booktitle = {Adjunct Proceedings of the 36th ACM Conference on Hypertext and Social Media},
pages = {25–29},
numpages = {5},
location = {
},
series = {HT Adjunct '25}
}

@inproceedings{Tricaud2024,
author = {Tricaud, Martin and Beaudouin-Lafon, Michel},
title = {Revisiting creative behaviour as an epistemic process: lessons from 12 computational artists and designers},
year = {2024},
isbn = {9798400717079},
publisher = {Association for Computing Machinery},
address = {New York, NY, USA},
url = {https://doi.org/10.1145/3638380.3638395},
doi = {10.1145/3638380.3638395},
booktitle = {Proceedings of the 35th Australian Computer-Human Interaction Conference},
pages = {175–190},
numpages = {16},
location = {Wellington, New Zealand},
series = {OzCHI '23}
}

\end{document}